\documentclass [aps, prx, reprint, floatfix, superscriptaddress]{revtex4-1}%
\usepackage{graphicx}
\usepackage{bm,amsmath,amssymb,mathrsfs,dcolumn}
\usepackage [colorlinks=true,linkcolor=blue,citecolor=blue,urlcolor=blue,linktoc=page,bookmarks=false,pdfstartview={FitH},pdfborder={0 0 0.0 [3 3]}]{hyperref}
\usepackage [usenames]{xcolor}
\hypersetup{pdfborder=0 0 0,colorlinks=true,citecolor=blue,linkcolor=blue}
\bibpunct{[}{]}{,}{n}{}{}
\begin{document}
\title{Interfacial spin Hall effect and spin swapping in Fe$\mid$Au bilayer from first principles}
\author {Song Li}
\affiliation {School of Science, Tianjin University, Tianjin 300072, China}
\affiliation {Department of Physics, Beijing Normal University, Beijing 100875, China}
\author {Ka Shen}
\email{kashen@bnu.edu.cn}
\affiliation {Department of Physics, Beijing Normal University, Beijing 100875, China}
\author {Ke Xia}
\affiliation {Department of Physics, Beijing Normal University, Beijing 100875, China}
\date{\today }

\begin{abstract}
  The interfaces in hybridized structures could give rise to rich phenomena, which open the way to novel devices with extraordinary performance. Here, we investigate the interface-related spin transport properties in Fe$\mid$Au bilayer based on first-principle calculation. We find that the spin Hall current in the Au side near the interface flows in the opposite direction to the bulk spin Hall current with the magnitude sensitive to the magnetization direction of Fe. This negative interfacial contribution is attributed to the spin dependent transmission within a few atomic layers, where a strong interfacial Rashba spin-orbit coupling exists. Surprisingly, the interfacial spin Hall currents are found to be not confined at the interface but extend tens of nanometers at low temperature, which is limited by momentum scattering and therefore much shorter than the spin diffusion length. In addition, the interfacial swapping spin currents, as a consequence of the spin precession under the interfacial Rashba field, are also obtained from our calculation and complete the full spin transport picture with all non-vanishing components. Our results suggest the importance of the interface engineering in spin-related transport in ferromagnetic$\mid$non-magnetic heterostructures and the possibility of manipulating the interfacial transport by the magnetization orientation of the magnetic layer.
\end{abstract}

\maketitle

\section{Introduction}
Spintronic devices have improved our experience in daily life with various outstanding applications in electronic products. In the meantime, more devices have been designed with the aims of lower energy consumption, more reliable storage and faster operation~\cite{stanciu:prb2006,binder:prb2006,kampfrath:nnano2013,jungwirth:nnano2016,marrows:science2016,Baierl:naturephotonics2016}. Seeking more efficient and controllable ways to generate spin currents is one of the main challenges in this field. To date, several techniques have been proposed and widely used in experiment, for instance, the electrical injection from ferromagnetic metal~\cite{Lou:natureP07,Tombros:nature07}, the spin pumping by magnetization procession~\cite{wei:nncomm2014,hou:apl2012,mosendz:prl2010,nakayama:prb2012}, the spin Seebeck effect with a temperature gradient across a magnetic insulator~\cite{Uchida:naturematerials2010,Jaworski:naturematerials2010,bauer:natruematerials2012,boona:ees2014}, and the spin Hall effect in heavy metals with strong spin-orbit coupling~\cite{sinova:reviewsofmodernphysics2015}. As most of them generate spin currents at the interface between two adjacent ferromagnetic and non-magnetic materials, the spin Hall effect is usually regarded as a bulk effect.

Recently, an interfacial contribution to the spin Hall effect was also demonstrated to exist and even be able to dominate the bulk spin Hall effect, which suggests the importance of taking into account the interfacial contribution in the analysis of the spin-Hall-related experimental data in heterostructures~\cite{hou:apl2012,kim:naturematerials2013,wang:prl2016}. For example, from the thickness dependence of the inverse spin Hall voltage across the bismuth film in Py$\mid$Bi bilayer structure, Hou et al.~\cite{hou:apl2012} found an interfacial term with the effective spin Hall angle of opposite sign to the bulk Bi. Similar phenomenon was found by Kim et al.~\cite{kim:naturematerials2013} who directly observed a sign change in the thickness dependence of the spin torques in Ta$\mid$CoFeB  bilayer and attributed the sign change to the competition between the bulk spin Hall effect in Ta and an effective term associated with interfacial Rashba spin-orbit coupling. Theoretically, Wang et al. \cite{wang:prl2016} recently reported their finding from first principles, where the Py$\mid$Pt interface enhances the effective spin Hall angle of Pt by one order of magnitude, however without sign change. It is therefore intuitively to ask why and how does the interface modify the magnitude and even reverse the sign of the spin Hall currents.

One simplified model to explore the interfacial spin-charge conversion is based on the analogy of two dimensional electron gas with Rashba spin-orbit coupling~\cite{Shen2014}. The injected spins in this model are assumed to accumulate in a relatively narrow area near the interface and produce a lateral charge current via the so-called spin galvanic effect or inverse Rashba-Edelstein effect~\cite{Ganichev:nature02,sanchez:nncomm2013}, which shares the same orthogonal relation between the spin and change flows as that in the spin Hall scenario and resembles the interfacial spin Hall effect. In such a two dimensional model, the interfacial electronic states are regarded to be isolated from the bulk states, which means that all possible consequences of the mixing between the interfacial and bulk states are ignored, hence some information, such as the spin injection efficiency from interface into bulk and the corresponding spin injection length, are unavailable in this model. Alternatively, a three dimensional model with the transport direction normal to the interface involved has been studied~\cite{wang:prbs2013,borge:prb2014,Amin:prb2016,Amin:prb2016b,Kim:prb17,borge:prb2017,Amin:arXiv2018}, where a Rashba-type interfacial spin-orbit coupling can be introduced by a voltage drop across the interface. Borge et al.~\cite{borge:prb2017} showed that when the electrons go through the interface between two metals, the normal spin-charge conversion due to the spin-dependent interfacial scattering and the anomalous one originating from the anomalous velocity of momentum-dependent spin-orbit coupling both can generate transverse currents but with opposite direction. The magnitude and the direction of the overall currents therefore depend on the competition between these two mechanisms and controllable by the the potential drop via the interface. Moreover, the spin swapping~\cite{Lifshits:prb2009,Shen:prb2014,Shen:prb2015,Saidaoui:prb2015,Espedal:prb2017,Pauyac:prl2018}, referring to the conversion between spin current flows due to spin-orbit-coupling-induced spin procession, originally predicted in bulk was also found to contain an interfacial piece~\cite{borge:prb2017}. The applicability of such toy models in real system still requires examination, which is one of our goals for the present paper.

\begin{figure} [t]
  \includegraphics [width=8.cm]{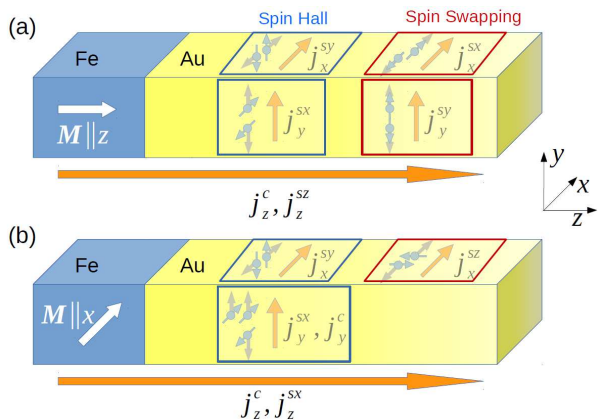}
  \caption{(Color online)
 Schematic of Fe$\mid$Au bilayer for the magnetization of Fe (a) parallel and (b) perpendicular to the transport direction. The orange arrows give all non-vanishing spin and charge currents in each configuration. The blue and red boxes correspond to the (inverse) spin Hall effect and spin swapping, respectively, where the gray arrows give the transverse velocity of a particular spin polarization represented by blue arrows. }%
\label{fig1}%
\end{figure}

The Fe$\mid$Au heterostructure, one potential candidate for spin-charge conversion devices, has been demonstrated to be able to generate Thz electromagnetic pulse via ultrafast laser exposure on the Fe side~\cite{kampfrath:nnano2013}. For the configuration with magnetization of Fe parallel to the interface, a spin current pulse generated by the ultrafast laser is injected into Au, where it converts into a time-dependent charge current due to inverse spin Hall effect and produces a THz radiation pulse. The generation efficiency of the THz pulse relies on the spin Hall angle, of which the interfacial effects discussion above may play a role but is out of consideration so far. Therefore, in the present work, we take Fe$\mid$Au bilayer as an example to investigate all bulk and interfacial spin-related transport properties based on first-principle calculation, including not only the (inverse) spin Hall effect but also the spin swapping effect in the presence of an applied current across the interface. We study two configurations with the magnetization of Fe parallel and normal to the interface, respectively. Interestingly, we find in both cases that the interfacial spin Hall current generated near the Fe$\mid$Au interface flowing in the opposite direction with respect to the bulk spin Hall current and this interfacial component can extend to tens of nanometers away from the interface. We interpret the generation of this interfacial component as the consequence of the spin-dependent transmission through the interfacial Rashba spin-orbit potential by considering the fact that the maximal interfacial spin Hall angle locates at the boundary of the Rashba region, justified by the vanishing interface-induced modification in the layer resolved density of states. The value of the maximal interfacial spin Hall angle depends on the temperature as well as the magnetization orientation. The penetration length of this interfacial spin Hall current is found to be limited by the momentum relaxation, instead of the spin relaxation, and is therefore much shorter  than the spin diffusion length. Moreover, in contrast to the spin Hall current, the spin swapping current is largest in the first atomic Au layer and present a quick decay within the Rashba region. We interpret this feature by the spin precession of the injected spins in the Rashba spin-orbit field. Far away from the interface, the spin swapping current induced by the bulk disorders shows a longer decay length determined again by the momentum scattering length.

The paper is organized as follows. In Sec. II, we will clarify the two configurations under study with all relevant quantities and briefly introduce the numerical method. In Sec. III, we will present our numerical results in both configurations. Both (inverse) spin Hall effect and spin current swapping will be discussed in this section. A summary will be given in Sec. IV.

\section{Method}
Fig.~\ref{fig1} illustrates the two configurations of the Fe$\mid$Au bilayer and all non-vanishing transverse currents. In case (a) with magnetization normal to the interface, i.e., $\mathbf M\| \hat z$, the applied electric current density $j_z^{c}$ across the interface is spin polarized along $z$ direction, accompanying a spin current $j_z^{sz}$. The spin Hall current densities read
\begin{equation}
  j_i^{sj}=\Theta_{\rm SH}^{ijz} \epsilon_{ijz}j_z^{c}
  \label{SH1}
\end{equation}
with $\epsilon_{ijz}$ and $\Theta_{\rm SH}^{ijz}$ being the Levi-Civita antisymmetric tensor and local spin Hall angle, respectively. The spin Hall angle in Eq.~(\ref{SH1}) satisfies $\Theta_{\rm SH}^{xyz}=\Theta_{\rm SH}^{yxz}$, guaranteed by the rotation symmetry about the $z$ axis, leading to
\begin{equation}
  j_y^{sx}=-\Theta_{\rm SH}^{yxz}j_z^{c}=-j_x^{sy}.
  \label{SHa}
\end{equation}
In case (b) with the magnetization along $x$ direction, $\mathbf M\| \hat x$, the magnetization breaks the rotation symmetry near the interface. The effective interfacial spin Hall angle becomes anisotropic, i.e., $\Theta_{\rm SH}^{xyz}\ne\Theta_{\rm SH}^{yxz}$, hence
\begin{eqnarray}
  j_y^{sx}\ne -j_x^{sy}.
  \label{Sh2}
\end{eqnarray}
In the meanwhile, the inverse spin Hall effect partially converts the injected spin current density $j_z^{sx}$ of case (b) into a charge current density
\begin{equation}
  j_y^{c}=-\Theta_{\rm SH}^{yxz} j_z^{sx}.
  \label{SHb}
\end{equation}
according to $j_i^{c}=\Theta_{\rm SH}^{ijz} \epsilon_{ijz}j_z^{sj}$~\cite{Shen:prb2014}.

The swapping spin currents can be generally expressed by~\cite{Lifshits:prb2009,Shen:prb2014,Shen:prb2015}
\begin{equation}
  j_i^{sj}=\kappa([j_j^{si}]^{(0)}-\delta_{ij}[j_l^{sl}]^{(0)}),
  \label{sw}
\end{equation}
where $[j_j^{si}]^{(0)}$ stand for the injected primary spin currents, $j_z^{sz}$ and $j_z^{sx}$ for $\mathbf M\| \hat z$ and $\mathbf M\| \hat x$, respectively. According to Eq.~(\ref{sw}), there are two non-zero swapping spin current densities 
\begin{equation}
  j_x^{sx}=j_y^{sy}=-\kappa j_z^z
  \label{sw2}
\end{equation}
for $\mathbf M\|\hat z$ and one 
\begin{equation}
  j_x^{sz}=\kappa j_z^{sx}
  \label{sw3}
\end{equation}
for $\mathbf M\|\hat x$.

In order to carry out all these transverse current densities from first principles, the [001] direction of the Fe bcc lattice is set along the transport diction (z). In the lateral plane, we use a $5\times 5$ supercell with periodic boundary condition and discrete the Brillouin zone into a $64\times 64$ mesh. In order to match the Au fcc lattice with Fe, we rotate it around $z$ axis by $\pi/4$ and stretch the Fe lattice constant by 0.6\%. The lattice constant of Au remains its nature value $a_{\rm Au}=4.0872${\AA}. To minimize the artificial effect of lattice distortion, we perform structure relaxation before transport calculation. The temperature effects are introduced in the manner of static phonon by a series of random atomic displacements with a temperature dependent root mean square. In the calculation, we average the outputs over tens static phonon configurations to reach the convergence. More technical details can be found in Refs.~\cite{wang:prb2008,Zhao:jap2011,wang:prl2016}. 

\begin{figure} [t]
\includegraphics [width=6cm]{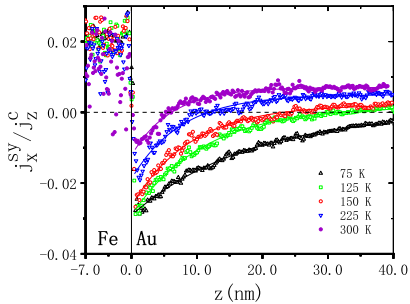}\caption{(Color online)
Spatial profile of the local spin Hall current density $j_x^{sy}$ normalized by the longitudinal current density $j_z^c$ at 75~K (black upper triangles), 125~K (green open squares), 150~K (red circles), 225~K (blue lower triangles), and 300~K (purple solid dots) with $\mathbf M\| \hat z$. The solid curves are fitting with Eq.~(\ref{fitting}).}%
\label{fig2}%
\end{figure}

\section{Results}
\subsection{Magnetization parallel to the transport direction ($\mathbf M\|\hat z$)}
We first take the magnetization of Fe along transport direction, i.e., $\mathbf M\| \hat z$. The spatial profile of the effective local spin Hall angle, i.e., the ratio between the local spin Hall current density $j_x^{sy}$ and the longitudinal charge current density $j_z^c$, is plotted in Fig.~\ref{fig2}. The other spin Hall current $j_y^{sx}$  is omitted because of the similarity [see Eq.~(\ref{SHa})]. Clearly, the effective spin Hall angle around the interface differs significantly from the bulk value not only the magnitude but also the sign. Specifically, three regimes can be recognized according to the different behaviors: (i) establishing the interfacial contribution within a common scale ($\sim 2$~nm); (ii) dissipating the interfacial part within a temperature-dependent length scale; (iii) saturating at the bulk spin Hall angle. At room temperature, the bulk spin Hall angle reads $\sim0.64$\%, which lies in the reported experimental range from 0.035\% to 11.3\%~\cite{sinova:reviewsofmodernphysics2015}. Theoretically, the intrinsic spin Hall conductivity was previously calculated from Berry phase formalism~\cite{Guo:jap2009}, which divided by the experimental value of longitudinal conductance~\cite{kittel1996introduction} gives a spin Hall angle around 0.16\% at room temperature, much smaller than ours. This might reflect the importance of the phonon-induced extrinsic mechanisms~\cite{Seki:naturematerials2008,skewtemgoriniprl:2015}, which are included in our calculation.

\begin{figure} [t]
\includegraphics [width=6cm]{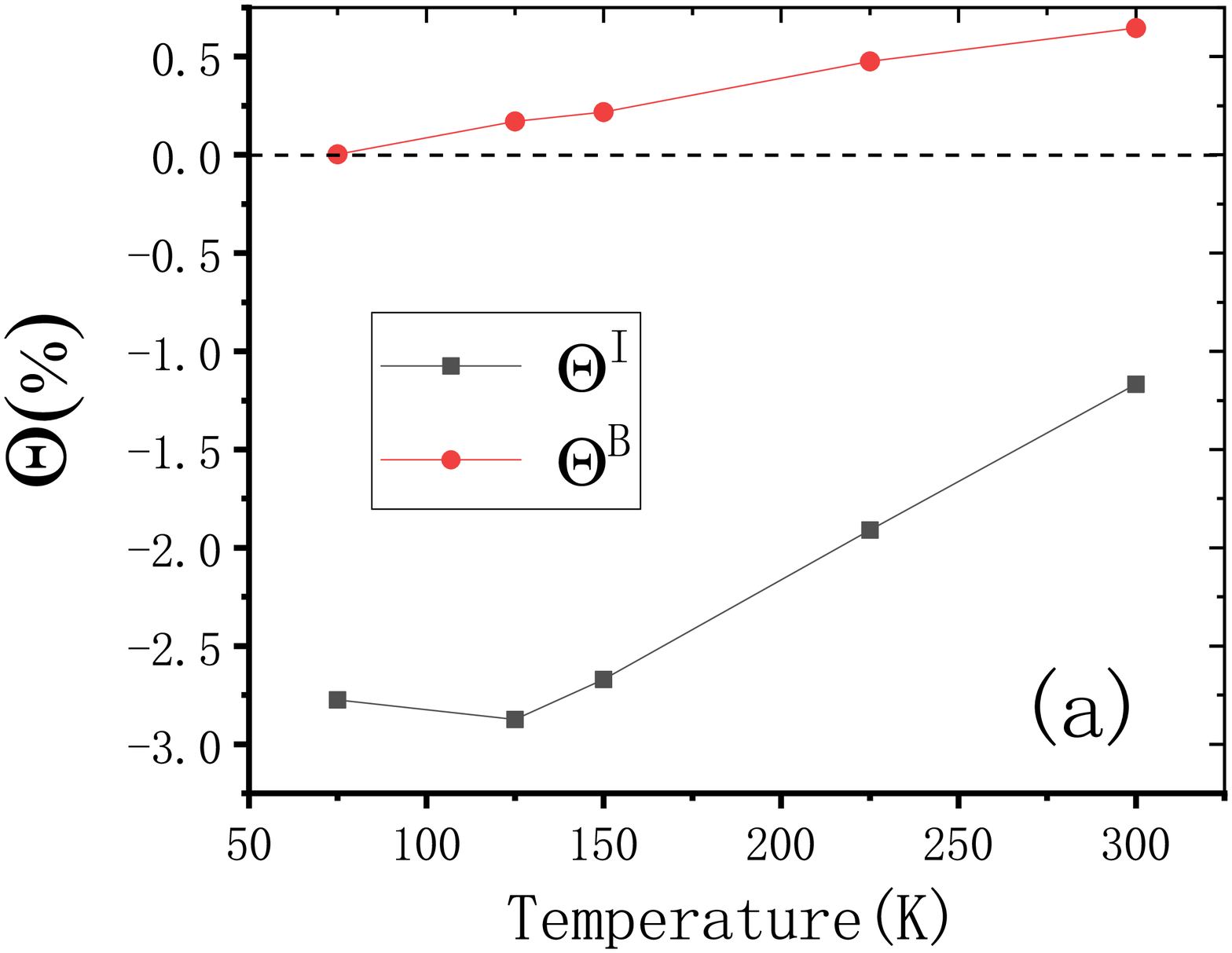}
\includegraphics [width=6cm]{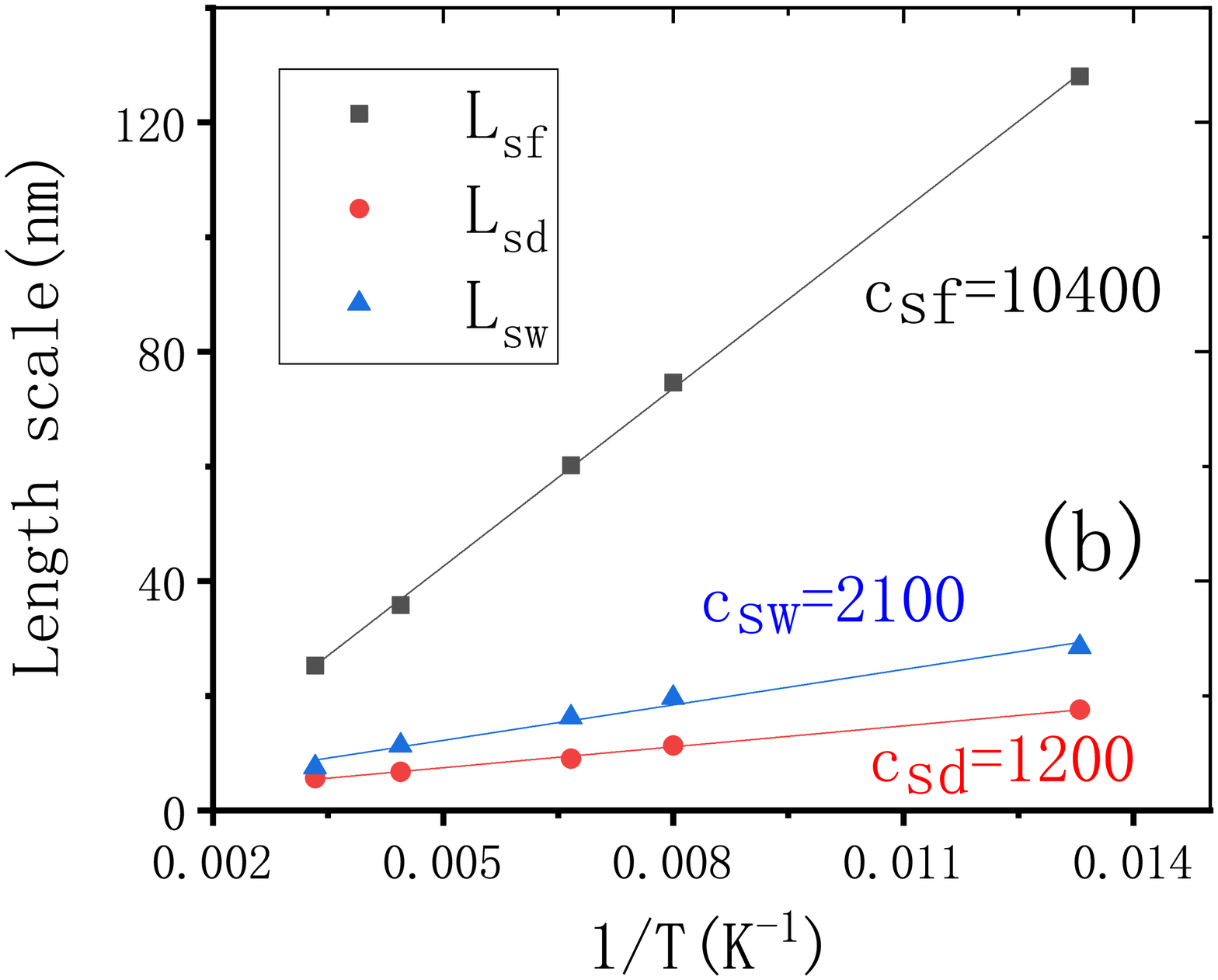}
\caption{(Color online)
 (a)  Temperature dependence of interfacial spin Hall angle (black squares) and bulk spin Hall angle (red dots). (b)Temperature dependence of spin diffusion length (black squares), the decay length of the interfacial spin Hall current (red dots) and spin swapping current (blue triangles). The solid lines in (b) correspond to fitting with $c_{i}/T$.
}%
\label{fig3}%
\end{figure}

As shown in Fig.~\ref{fig2}, the interfacial contribution at room temperature extends to $10\mu$m, much longer than that found in Pt~\cite{wang:prl2016}. When the temperature decreases, the interfacial part covers even a longer distance. Assuming an exponential decay of the interfacial term, we fit the results ($z>2$~nm) in Fig.~\ref{fig2} by
\begin{equation}
  (j^{sy}_x/j^c_z)(z)=\Theta^{\rm B}+\Theta^{\rm I}e^{-z/L_{\rm sd}},
\label{fitting}
\end{equation}
to extract the decay length $L_{\rm sd}$ as well as the bulk and interfacial spin Hall angles, $\Theta^{\rm B}$ and $\Theta^{\rm I}$. The outputs are summarized in Fig.~\ref{fig3}. Notice that although the local spin all angle at low temperatures does not saturate to the bulk value at the longest distance ($40\mu$m) of our computational ability, the bulk spin Hall angle can still be obtained from the fitting. Fig.~\ref{fig3}(a) shows that $\Theta^{\rm B}$ increases linearly with increasing temperature, suggesting the dominant role of the phonon skew scattering mechanism~\cite{skewtemgoriniprl:2015,wang:prl2016}. The interfacial contribution is typically of much larger magnitude than the bulk value and its temperature dependence is opposite to the bulk term, namely instead of decreasing, it increases at lower temperature. By taking into account the increasing decay length $L_{\rm sd}$ shown in Fig.~\ref{fig3}(b), we expect a more remarkable interfacial spin Hall contribution at extremely low temperature, which is confirmed by a zero-temperature calculation, where the local spin Hall angle is almost a constant ($3\%$) across the entire Au layer (not shown).

\begin{figure} [t]
\includegraphics [width=8.5cm]{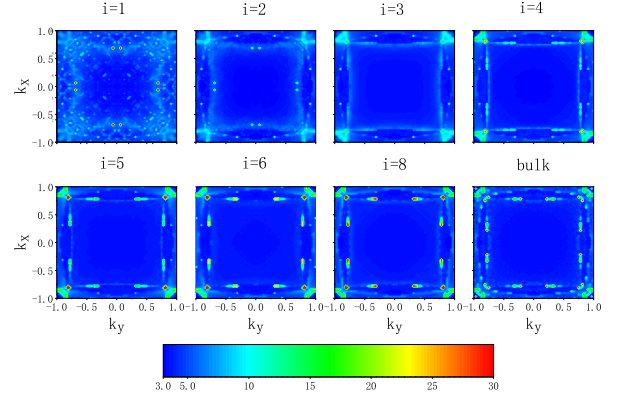}\caption{(Color online)
    Layer resolved density of states for layer index i=1, 2, 3, 4, 5, 6, 8 and bulk mapped in lateral Brillouin zone with $\mathbf M\|\hat z$. The wave vectors are in unit of $\pi/a$, where a is the effective lattice constant of the lateral supercell.}%
\label{fig4}%
\end{figure}

To understand the microscopic origin of the interfacial spin Hall effect, we project the layer resolved density of states near the interface into the reciprocal space in Fig.~\ref{fig4}, where for comparison, the bulk density of states (a single atomic layer far away from the interface) is also given. As one can see that only the very first couple of layers are strongly modified by the interface, which is in the comparable length scale with the establishing regime (i) defined above. This implies that the interfacial spin Hall current may be induced by the spin-dependent interfacial potential and its long-distance-living feature may reflect the propagating property. The generation process can be understood as a consequence of spin-dependent transmission. We naively assume a Rashba-type spin orbit coupling due to the potential mismatch at the interface with its strength strongly depending on the distance to the interface,  i.e.,
\begin{equation}
H_{\rm Rashba}=\alpha(z)(p_x\sigma_y-p_y\sigma_x),
\label{rashba}
\end{equation}
where $\alpha(z)\propto \partial_z V(z)$ with $V(z)$ being the potential drop or lift at the interface. Assuming $\alpha(z)>0$, the electronic states with a net polarization in $y$ direction, $\langle \sigma_y \rangle >0$, experience a potential lift (drop) if its lateral momentum component $p_x>0$ ($ p_x <0$) according to Eq.~(\ref{rashba}). The corresponding transmissions with $p_x<0$ is therefore larger than those with $p_x>0$. In other words, the transmitted spin species with $\langle \sigma_y \rangle >0$ harvest a net momentum along $-x$ direction, leading to a net spin polarized current $j_{x}^{sy}<0$. Similarly, for the other spin species with $\langle \sigma_y \rangle <0$ the transmission of $p_x>0$ states is larger, again giving $j_x^{sy}<0$. Since the accompanying charge currents of these two flows compensates with each other, a net lateral pure spin current is injected into Au. With the same reason, $j_y^{sx}>0$ is simultaneously generated. This spin-dependent transmission picture is also consistent with the fact that in the Fe side in Fig.~\ref{fig2}, the spin Hall current has opposite sign to the Au side near interface, because the total spin (current) of the reflected and transmitted beams should conserve once the spin flipping at the interface is negligibly small. Such an interfacial scattering has also been discussed by Amin et al.~\cite{Amin:prb2016b,Amin:prb2016b,Amin:arXiv2018} in the presence of a lateral electric current. Notice that the spin current reverses its direction, i.e., $j_x^{sy}>0$, if $\alpha(z)$ changes sign. This means that both positive and negative interfacial spin Hall angles are possible and in principle controllable by fabricating different heterostructures to tune the relative potential $V(z)$~\cite{borge:prb2017}.

Once the injected electrons propagate out of the establishing regime, the Rashba field disappears and the spin-momentum-locked scattering is suppressed. In this case, even without any spin-flip process, the momentum scattering can redistribute the two spin species in momentum space. As a result, each spin species lose net lateral velocity and the interface-induced pure spin Hall currents are dissipated. This process is therefore governed by momentum relaxation length or mean free path, which grows at lower temperature as shown in Fig.~\ref{fig3}(b). Intuitively, this length scale should also manifest itself in the bulk spin Hall current, especially near the interfaces or the boundaries. 

With all these understandings, we write out the general form of the spin Hall current density as
\begin{equation}
  j_x^{sy}(z)=j_z^{c}\int_0^{z} \frac{d z'}{L_{\rm sd}} [\Theta_{\rm SH}^{\rm B}+\Theta_{\rm SH}^{\rm I}(z')] e^{-(z-z')/L_{\rm sd}},
  \label{sh_mix}
\end{equation}
where the exponent $\exp[-(z-z')/L_{\rm sd}]$ is introduced to catch the damping of the generated spin Hall current during its propagation along $z$ direction. Here, the bulk spin Hall angle $\Theta_{\rm SH}^{\rm B}$ is a constant while the interfacial part $\Theta_{\rm SH}^{\rm I}(z')=\Theta_{\rm SH}^{\rm I}\exp(-z'/L_{\rm I})$ with the establishing length $L_{\rm I}\simeq 2$~nm. By integrating Eq.~(\ref{sh_mix}), we obtain
\begin{eqnarray}
  (j_x^{sy}/j_z^{c})(z)&=&\Theta_{\rm SH}^{\rm B}-\Theta_{\rm SH}^{\rm I}({L_{\rm sd}}/{L_{\rm I}}-1)^{-1}e^{-z/L_{\rm I}}\nonumber\\
  &&\hspace{-0.5cm}+[\Theta_{\rm SH}^{\rm I}({L_{\rm sd}}/{L_{\rm I}}-1)^{-1}-\Theta_{\rm SH}^{\rm B}]e^{-z/L_{\rm sd}},
  \label{sh_mixb}
\end{eqnarray}
where the second term corresponds to the quick establishment of the interfacial contribution. The other two terms are those in Eq.~(\ref{fitting}) with fitting parameter expressed as $\Theta^{\rm B}=\Theta^{\rm B}_{\rm SH}$ and $\Theta^{\rm I}=\Theta^{\rm I}_{\rm SH}(L_{\rm sd}/L_{\rm I}-1)^{-1}-\Theta^{\rm B}_{\rm SH}$. The decreases of $\Theta^{\rm I}$ at high temperature can be interpreted by the disorder-induced destruction of the Rashba potential.

\begin{figure} [t]
  (a)\includegraphics [width=6cm]{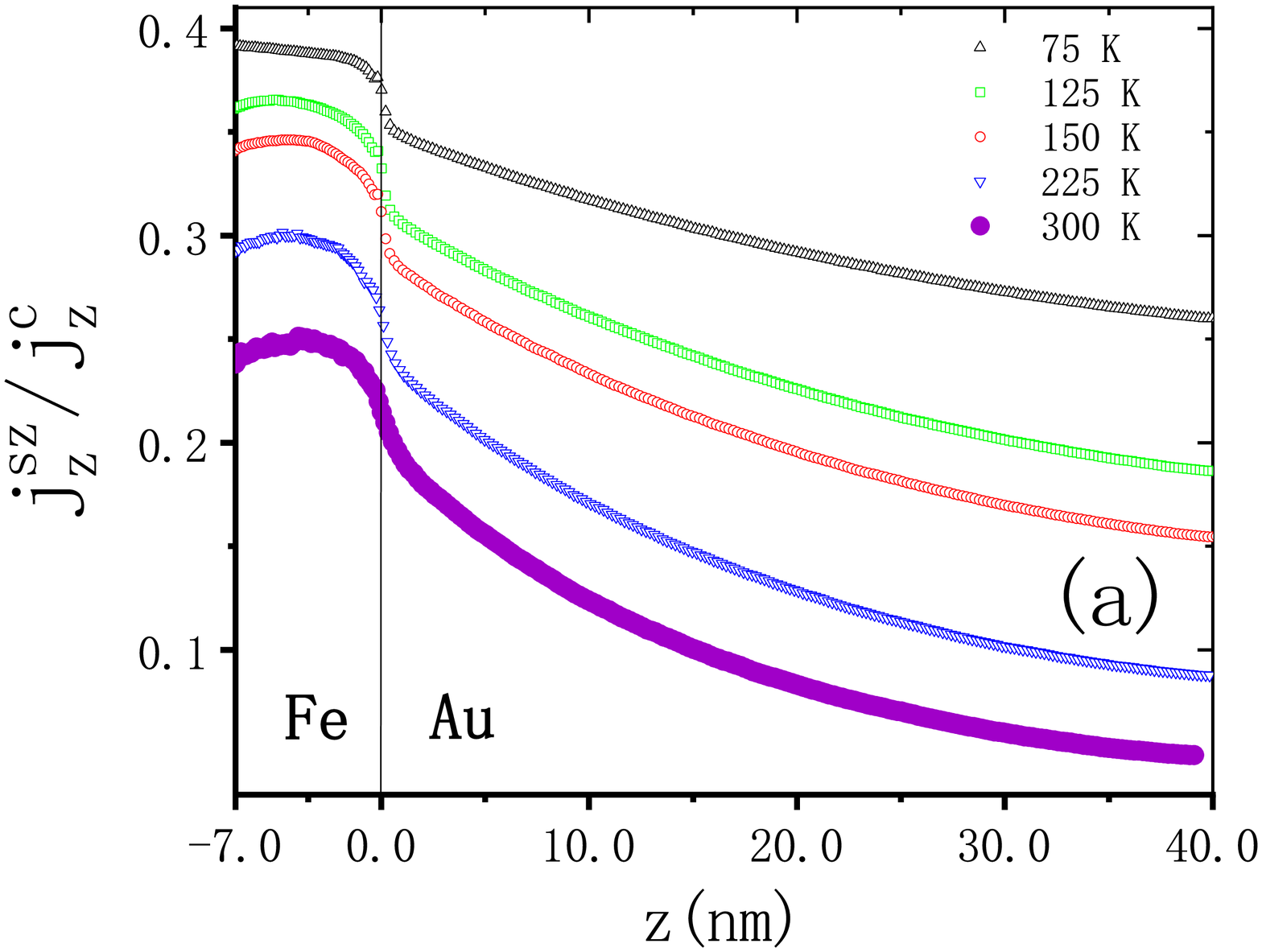}
  (b)\includegraphics [width=6cm]{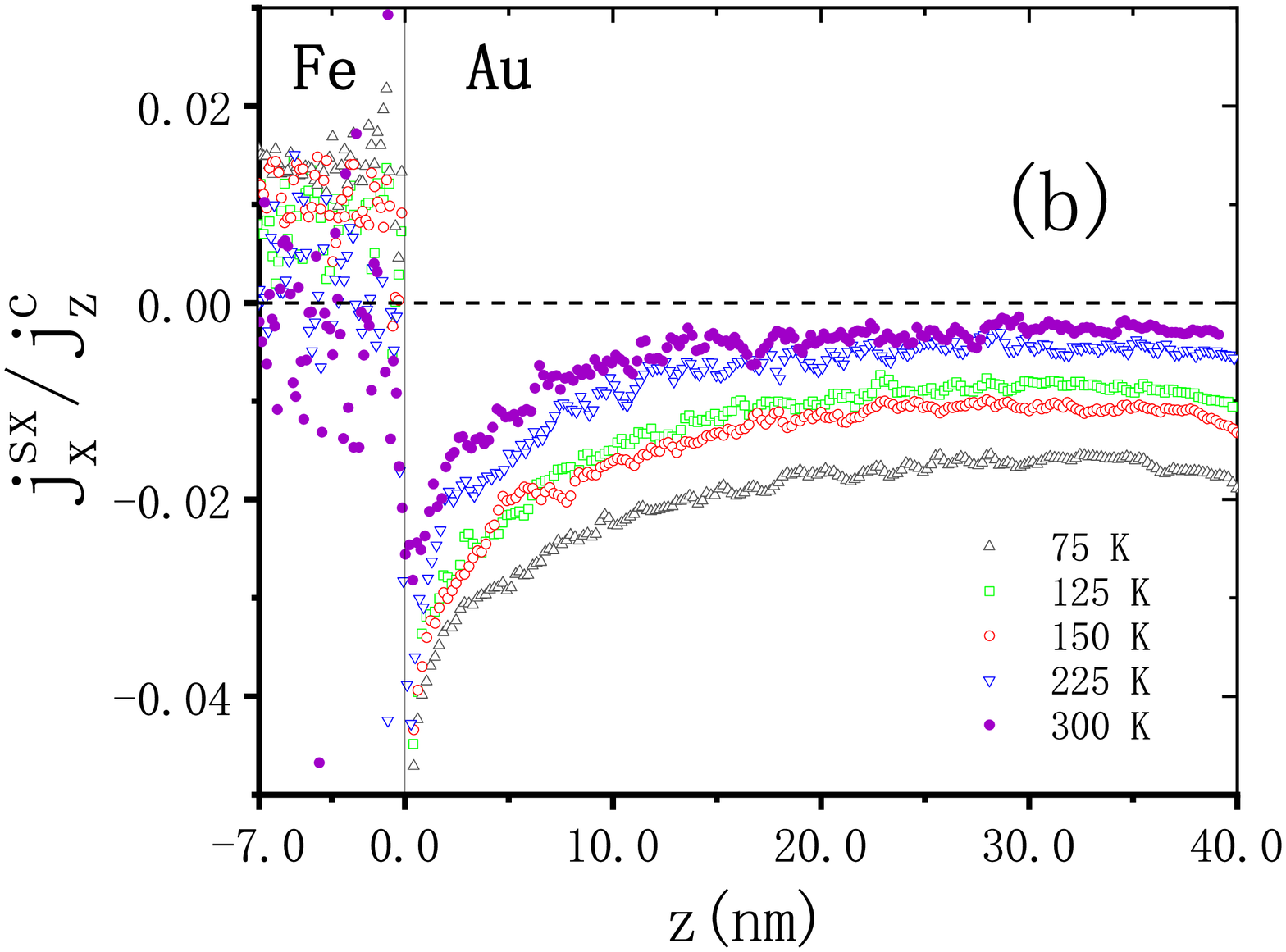}
  \caption{(Color online)
  Spatial profiles of the normalized (a) longitudinal spin current density $j_z^{sz}$ and (b) spin swapping current density for different temperatures with $\mathbf M\| \hat z$.}%
\label{fig5}%
\end{figure}

For comparison, in Fig.~\ref{fig5} (a), we plot the spatial profile of the local longitudinal spin current $j_z^{sz}$ from the same calculation, which also follows an exponential decay with the decay lengths plotted as $L_{\rm sf}$ in Fig.~\ref{fig3}(b). One can see that $L_{\rm sf}$ is systematically longer than $L_{\rm sd}$. By expressing $j_z^{sz}\propto v_z \langle \sigma_z \rangle$ and $j_z^c\propto v_z$ with drift velocity $v_z$, one obtains
\begin{equation}
(j_z^{sz}/j_z^c)(z)\propto \langle \sigma_z \rangle \propto e^{-z/L_{\rm sf}}.
\end{equation} 
Therefore, $L_{\rm sf}$ corresponds to the spin relaxation length, which is determined by Elliott-Yafet mechanism~\cite{Elliott54,Yafet63} according to its increase with weaker scattering at lower temperature. The spin diffusion length at room temperature is around 25~nm, comparable with the reported experimental values~\cite{sinova:reviewsofmodernphysics2015}. One may also notice a sudden drop of spin current $j_z^{sz}$ from Fig.~\ref{fig5}(a) in the establishing regime. This so called interfacial spin memory loss is related to the spin-flip scattering due to the interface spin-orbit coupling~\cite{Amin:prb2016,Amin:prb2016b}  as well as the spin swapping discussed below.

Fig.~\ref{fig5}(b) shows the spatial distribution of the swapping spin current density $j_x^{sx}$ (similar for $j_y^{sy}$) generated by the primary spin current density $j_z^{sz}$. In contrast to the spin Hall currents, here the spin swapping current does not change sign. More importantly, the spin swapping current is largest at the first atomic Au layer and shows a quick decay within a length scale comparable to the drop of $j_z^{sz}$ in Fig.~\ref{fig5}(a), indicating that the interfacial spin swapping may be responsible to the spin memory loss at interface via partially converting $j_z^{sz}$ to $j_x^{sx}$~\cite{borge:prb2017}. The microscopic process can be understood from the spin precession in the interfacial Rashba field~\cite{borge:prb2017}. Describing the injected spin density polarized along $z$ direction by density matrix $\Delta\rho={\cal S}_z(z) \sigma_z$, from the steady state solution of the equations of motion $i[H_{\rm Rashba},\rho]=-(\rho-\rho_0)/\tau$, one can estimate the correction in the density matrix due to momentum-dependent spin precession as
\begin{eqnarray}
  \Delta\rho'(x)&\simeq&\tau {\cal S}_z(z)\alpha(z) (p_x \sigma_x+p_y \sigma_y),
\label{drho1}
\end{eqnarray} 
which apparently supplies contributions to $j_x^{sx}(z)$ and $j_y^{sy}(z)$. Here, $\tau$ and ${\cal S}_z$ are the momentum scattering time and the local spin density, respectively. According to Eq.~(\ref{drho1}), the strength of the interfacial spin swapping is proportional to the product of the local Rashba coefficient $\alpha(z)$ and local spin density ${\cal S}(z)$.  For the injected spins with a group velocity almost transverse $p_z\ll |p_{x,y}|$, they can stay near the interface for a relative long time. In this sense, they are partially ``localized'' and form a quasi-two dimensional electron gas, which is confirmed by the larger density of states at the interface compared to bulk Au in the calculation (not shown). Note that the potential mismatch may also help to form the quasi-two dimensional electron gas. As both spin density ${\cal S}_z$ and Rashba coefficient $\alpha$ rapidly decrease with increasing the distance, the swapping spin current shows a sharp decrease within the first few layers. Such a spin precession simultaneously reduces the total spin polarization along $z$ direction, resulting in a spin memory loss.

Outside the Rashba region, the spin swapping efficiency, $\kappa$, in Eq.~(\ref{sw2}) becomes a constant. By taking into account the transport of the generated swapping current as done for spin Hall effect in Eq.~(\ref{sh_mix}), we express the spin swapping current ($z> L_{\rm I}$)
\begin{equation}
  j_x^{sx}(z)\simeq  -J^{\rm I}e^{-z/L_{\rm sd}'}-\kappa \int_0^{z} \frac{d z'}{L_{\rm sd}'} j_z^{sz}(z') e^{-(z-z')/L_{\rm sd}'},
  \label{swapping_c}
\end{equation}
with $J^{\rm I}$ representing the propagating part of the interfacial contribution. Since its dissipation is also caused by momentum scatterings, the decay length is assumed to be the same as $L_{\rm sd}$ in the interfacial spin Hall current discussed above, i.e., $L_{\rm sd}'=L_{\rm sd}$. Substituting the exponential function for the primary spin current, $j_z^{sz}(z')\simeq (J^{\rm B}/\kappa)e^{-z'/L_{\rm sf}}$ into Eq.~(\ref{swapping_c}), we obtain
\begin{eqnarray}
%  j_x^{sx}(z)\simeq J^{\rm I}e^{-z/L_{\rm sd}} -J^{\rm B} (1-L_{\rm sd}/L_{\rm sf})^{-1}(e^{-z/L_{\rm sf}}-e^{-z/L_{\rm sd}}),
  j_x^{sx}(z)&\simeq&  -J^{\rm B} (1-L_{\rm sd}/L_{\rm sf})^{-1}e^{-z/L_{\rm sf}}\nonumber\\
  &&-[J^{\rm I}-J^{\rm B} (1-L_{\rm sd}/L_{\rm sf})^{-1}]e^{-z/L_{\rm sd}},
  \label{swapping}
\end{eqnarray}
where both length scales are involved. It turns out that when the interfacial term is stronger than the bulk one, i.e., $J^{\rm I}>J^{\rm B}$, and $L_{\rm sd}\ll L_{\rm sf}$, Eq.~(\ref{swapping}) reduces to 
\begin{equation}
  j_x^{sx}(z) \simeq -J^{\rm I} e^{-z/L_{\rm sd}} -J^{\rm B}e^{-z/L_{\rm sf}},
\end{equation}
where the first term dominates. Therefore, the decay length $L_{\rm sw}$ extracted by using single exponential function to fit the results (2~nm$<z<20$~nm) in Fig.~\ref{fig5}(b) is only slightly longer than  $L_{\rm sd}$ [see Fig.~\ref{fig3}(b)]. The non-monotonic feature in $z>30$~nm might be related to the reflection at the interface ($z=40$~nm) with the right lead.

\begin{figure} [t]
\includegraphics [width=8.5cm]{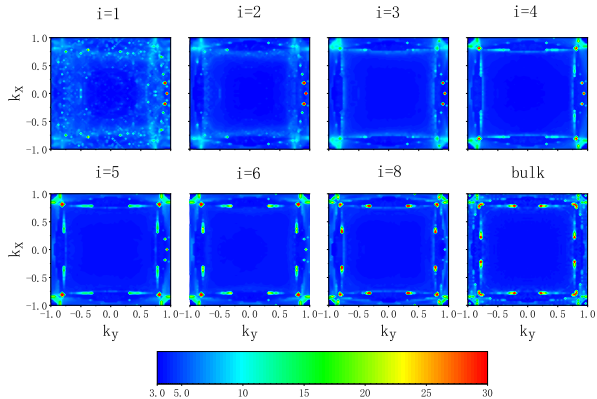}\caption{(Color online)
    Layer resolved density of states mapped in lateral Brillouin zone for $\mathbf M\| \hat x$.}%
\label{fig6}%
\end{figure}

\subsection{Magnetization perpendicular to the transport direction ($\mathbf M\|\hat x$)}
Now we turn to the other configuration with the magnetization of Fe perpendicular to the transport direction, i.e., $\mathbf M\|\hat x$. As discussed in Method section, the magnitudes of $j_y^{sx}$ and $j_x^{sy}$ in this case do not have to be equal, which could also be expected from the layer resolved density of states shown in Fig.~\ref{fig6}, where four-fold rotation symmetry in Fig.~\ref{fig4} reduces to a mirror symmetry with respect to  $y$ axis. Since this configuration is widely used in spin-charge conversion experiment,  it is important to check the influence of the magnetization orientation.

\begin{figure} [t]
(a)\includegraphics [width=6cm]{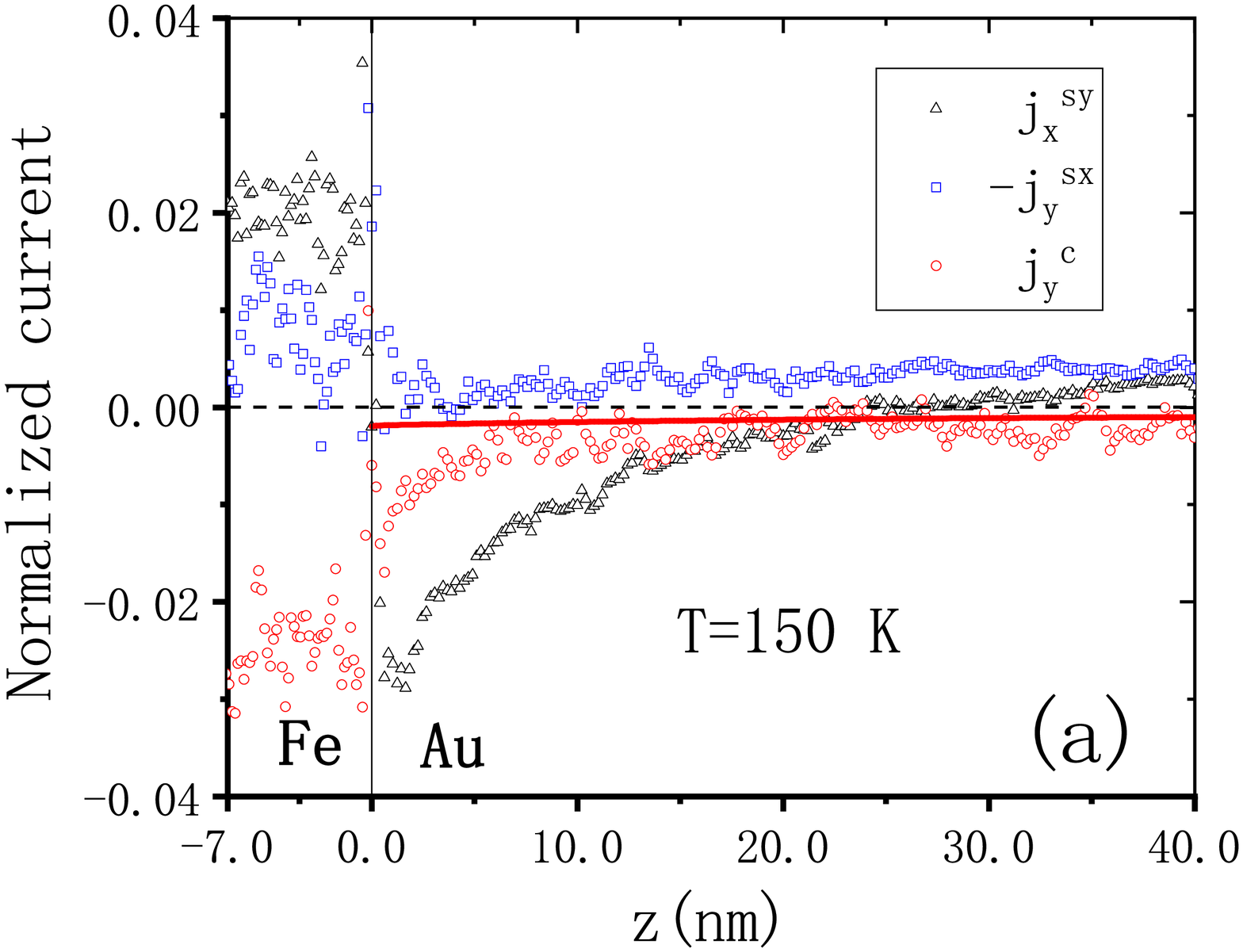}
(b)\includegraphics [width=6cm]{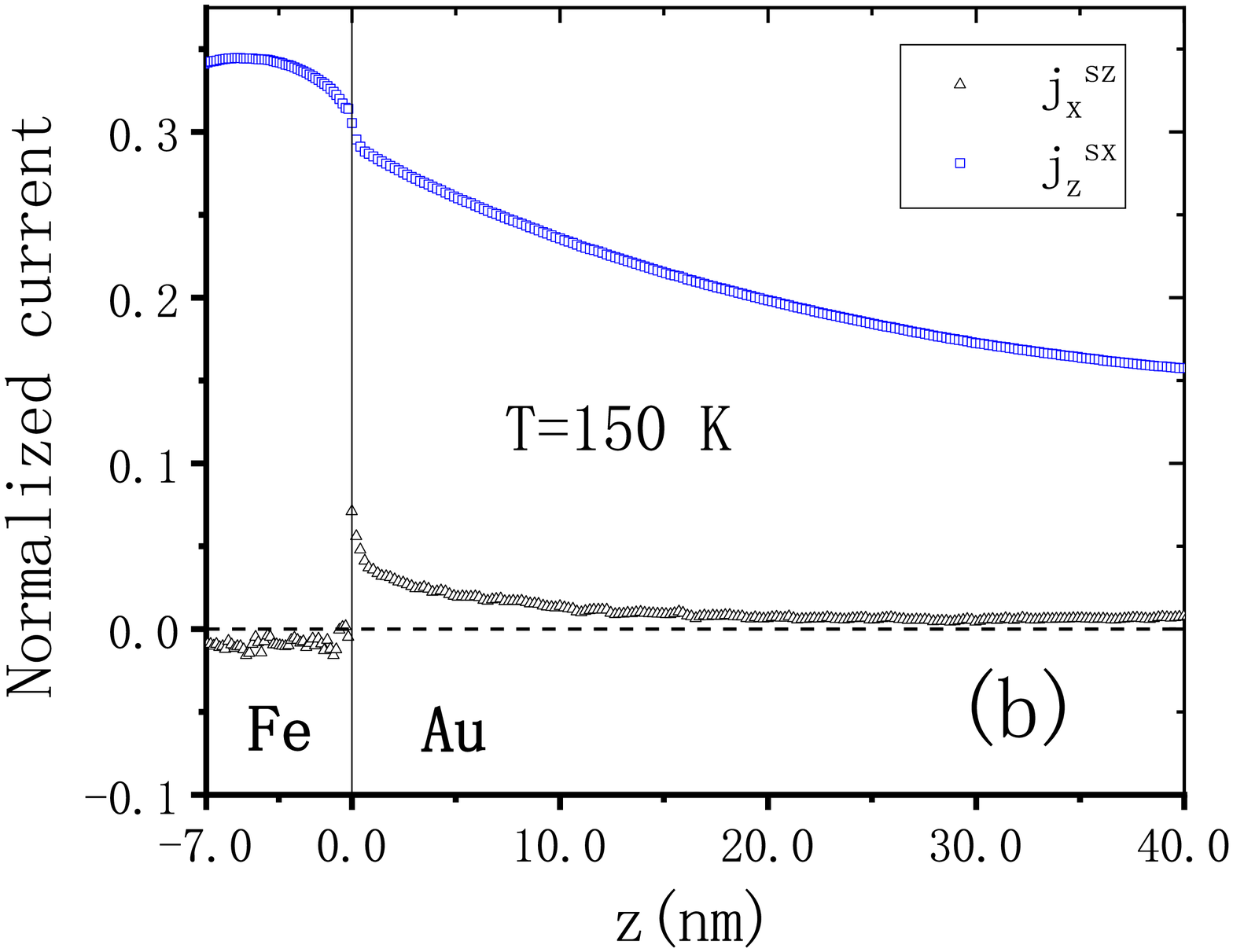}
\caption{(Color online)
  (a) Spatial profiles of the local transverse spin Hall current densities, $j_x^{sy}$ and $j_y^{sz}$, and the local transverse charge current density $j_y^c$ normalized by the longitudinal current density $j_z^c$ with $\mathbf M\|\hat x$ at 150K. (b) Spatial profiles of the primary spin current density $j_z^{sx}$ and swapping current $j_x^{sz}$. }%
\label{fig7}%
\end{figure}

Fig.~\ref{fig7} shows the spatial distributions of the two transverse spin Hall current densities at 150~K. While they approach each other and saturate to the same value (bulk spin Hall angle), they however differ significantly near the interface. Specifically, $j_x^{sy}$ remains very similar to the previous case shown in Fig.~\ref{fig2}, because of the fact that $j_x^{sy}$ is related to the spin polarization in $y$ direction, which is insensitive to the rotation of the magnetization around $y$ axis. The interfacial contribution in $j_y^{sx}$ becomes much smaller. One can express $j_y^{sx}$ by
\begin{equation}
  j_y^{sx} \simeq \sum_{\mathbf p}|p_y|[(\Gamma^{\uparrow}_{p_y>0}-\Gamma^{\uparrow}_{p_y<0}){\cal S}_x^\uparrow-(\Gamma^{\downarrow}_{p_y>0}-\Gamma^{\downarrow}_{p_y<0}){\cal S}_x^\downarrow],
\label{jyx} 
\end{equation} 
where ${\cal S}_x^\uparrow$ and ${\cal S}_x^\downarrow$ stand for the spin-up and spin-down densities of states, separately, with respect to the $x$-axis. Assuming $\alpha(z)>0$ in Eq.~(\ref{rashba}), 
\begin{eqnarray}
\Gamma^\uparrow_{p_y>0}&\sim&\Gamma^\downarrow_{p_y<0}\sim\Gamma_0+\delta\label{gm1},\\
\Gamma^\downarrow_{p_y>0}&\sim&\Gamma^\uparrow_{p_y<0}\sim\Gamma_0-\delta\label{gm2},
\end{eqnarray}
with $\Gamma_0$ and $\delta$ being spin-independent and dependent parts of the transmission. By substituting Eqs.~(\ref{gm1}) and (\ref{gm2}) into Eq.~(\ref{jyx}), we obtain
\begin{equation}
  j_y^{sx}\sim 2\delta\sum_{\mathbf p}|p_y| (S^x_\uparrow+S^x_\downarrow),
\end{equation}
depending mainly on the total density of states. The accompanying charge current is given by
\begin{eqnarray}
  j_y^{c} &\simeq& \sum_{\mathbf p}|p_y|[(\Gamma^{\uparrow}_{p_y>0}-\Gamma^{\uparrow}_{p_y<0})S^x_\uparrow+(\Gamma^{\downarrow}_{p_y>0}-\Gamma^{\downarrow}_{p_y<0})S^x_\downarrow]\nonumber\\
  &\sim&2\delta\sum_{\mathbf p}|p_y| (S^x_\uparrow-S^x_\downarrow),
\label{jyc} 
\end{eqnarray}
proportional to the spin polarization as well as the parameter $\delta$. For the case with the magnetization perpendicular to the $x$ direction, for instance, the previous case with $\mathbf M\| \hat z$, $S^x_\uparrow=S^x_\downarrow$, therefore $j_y^c$ vanishes and $j_y^{sx}$ survives.

 We should point out that the difference between the two transmission coefficients, $\delta$, depends on the magnitude of the Rashba-induced potential. In our case, the Rashba potential is relatively weak, therefore, $\delta$ is sensitive to the magnetization direction and is suppressed significantly when the magnetization lies in $x$ direction. In the mean time, the interfacial contribution to inverse spin Hall current $j_y^c$ is also suppressed according to Eq.~(\ref{jyc}), and the bulk spin Hall mechanism dominates. This picture is consistent with the numerical results shown in Fig.~\ref{fig7}(a), where no sign change is seen in the inverse spin Hall current (red circles). One may thus expect the transverse charge current $j_y^c$ comparable to the product of $j_x^{sz}$ and the bulk spin Hall angle $\Theta^B=0.2\%$ as plotted by the red solid curve, which however does not agree with the numerical results near the interface. The deviation actually reveals the transmission of the anomalous Hall current from Fe (see the $z<0$ region). For a large Rashba potential, $\delta$ is robust against the change of magnetization direction as shown in Py$\mid$Pt, where a huge interfacial spin Hall current is generated by strong Rashba potential and the magnitude of the interfacial $j_y^{sx}$ is only slightly smaller than that of $j_x^{sy}$ for $\mathbf M\| \hat x$~\cite{wang:prl2016}. 

The spin swapping current and the primary spin current are plotted in Fig.~\ref{fig7}(b). Both are very similar to those in the previous case with $\mathbf M\| \hat z$. The sign of the swapping current here is positive as expected by Eq.~(\ref{sw3}). Following the derivation above, the interfacial spin procession, in the present case of $\mathbf M\| \hat x$, causes a correction in density matrix
\begin{eqnarray}
  \Delta\rho'(x)&\simeq&-\tau {\cal S}_x(z)\alpha(z) p_x \sigma_z,
\label{drho2}
\end{eqnarray} 
which leads to a large interfacial $j_x^{sz}$ with opposite sign to $j_z^{sx}$ in Eq.~(\ref{drho1}).

% depends on the magnitude of the Rashba-induced potential and the longitudinal kinetic energy ($p^2_z/2m$) of the incident electrons. Intuitively, if the longitudinal kinetic is much larger than the Rashba term, the potential modification due to Rashba field is negligible, leading to a reduction in $\delta$ and the transmitted spin Hall current. Following this argument, we suppose that the distinct magnitudes of $j_y^{sx}$ and $j_x^{sy}$ results from the anisotropic dispersion of Fe. The magnetization sensitivity of the interfacial spin Hall effect allows the manipulation by controlling the magnetization orientation.

\section{Summary and discussion}
In summary, we employ first-principle method to study the spin transport properties in Fe$\mid$Au bilayer system in the presence of an applied electric current across the interface and obtain strong interfacial contributions in both (inverse) spin Hall effect and spin swapping effect, which can be well-explained by the spin dependent transmission and spin precession under the interfacial Rashba spin-orbit field, respectively. We find that the negative interfacial term of the spin Hall currents is induced within several atomic layers near the interface, where the layer resolved density of states reveals a strong modification in the electronic band structure. Very interestingly, the decay length of the interfacial spin currents is not a constant but increases with decreasing the temperature and its value is typically much shorter than the spin diffusion length. This is because of the fact that it is mainly limited by momentum scatterings. We also show that the interfacial spin Hall term is controllable by tuning the magnetization direction of Fe and becomes anisotropic if the magnetization is away from the normal direction of the interface plane. For the spin swapping effect, we found that the maximal value of the spin swapping current occurs at the first atomic layer and a quick decay nearby no matter the magnetization lies in or normal to the interface, which is explained as the consequence of the spin precession of quasi-two dimensional near the interface. Our calculation indicates the importance of the interfacial effects in spin transport in bilayer or other heterostructure devices.

Finally, we discuss the experimental test of our predictions. One of the standard approaches to determine the spin Hall angle is to measure the inverse spin Hall voltage in the spin pumping experiment. However, according to Fig.~\ref{fig7}(a), the negative interfacial spin Hall contribution in Fe$\mid$Au bilayer is strongly suppressed in the typical spin pumping configuration with the magnetization exactly in-plane. In this sense, one may need a tilted configuration with the magnetization containing both in-plane and out-of-plane components. The predicted behaviors of the interfacial term thus can be tested from the thickness and temperature dependences of the inverse spin Hall voltage via spin pumping or spin Seebeck effect.

\begin{acknowledgements}
This work was financially supported by National Key Research and Development Program of China (Grant No. 2017YFA0303300) and the National Natural Science Foundation of China (No.61774017, No. 11734004, and No. 21421003). K.S. acknowledges the Recruitment Program of Global Youth Experts and the Fundamental Research Funds for the Central Universities (Grant No. 2018EYT02).
\end{acknowledgements}

\bibliography{ref}

\end{document}